\documentclass[twocolumn,showpacs,preprintnumbers,amsmath,amssymb]{revtex4-1}
\usepackage{amsfonts}
\usepackage{amsmath}
\usepackage{graphicx}
\usepackage{dcolumn}
\usepackage{bm}
\usepackage{overpic}
\usepackage{booktabs}
\usepackage{color}

\begin{document}
\title{Degree heterogeneity in spatial networks with total cost constraint}
\author{Weiping Liu}
\author{An Zeng}\email{an.zeng@unifr.ch}
\author{Yanbo Zhou}
\affiliation{Department of Physics, University of Fribourg, Chemin du Mus\'{e}e 3, CH-1700 Fribourg, Switzerland}
\date{\today}

\begin{abstract}
Recently, In [Phys. Rev. Lett. 104, 018701 (2010)] the authors studied a spatial network which is constructed
from a regular lattice by adding long-range edges (shortcuts) with
probability $P_{ij}\sim r_{ij}^{-\alpha}$, where $r_{ij}$ is the Manhattan
length of the long-range edges. The total
length of the additional edges is subject to a cost constraint ($\sum r=C$). These networks have fixed optimal exponent $\alpha$
for transportation (measured by the average shortest-path length). However, we observe that the degree in such spatial networks is homogenously distributed, which is far different from real networks such as airline systems. In this paper, we propose a method to introduce degree heterogeneity in spatial networks with total cost constraint. Results show that with degree heterogeneity the optimal exponent shifts to a smaller value and the average shortest-path length can further decrease. Moreover, we consider the synchronization on the spatial networks and related results are discussed. Our new model may better reproduce the features of many real transportation systems.
\end{abstract}
\maketitle

Spatial features play a significant role in the transportation networks~\cite{PNAS7794}, Internet~\cite{PNAS13382}, mobile phone networks~\cite{Science1071}, power grids~\cite{PRE025103}, social networks~\cite{PNAS11623} and neural networks~\cite{Nature393}. In the past decade, many systems have been modeled by complex networks where nodes and links are embedded in space. In these models, nodes are located in the plane and the geometric distance between nodes is well defined. Then links are constructed according to rules based on spatial indices~\cite{PRE6637102,PRL102238702,PRL102238703,NP481,EPJB63273,PRE70056122}. Growing spatial networks were also studied~\cite{PRE6626118,PRE75036106}. Generally, these model of spatial networks are able to reproduce some properties of the real systems such as community structure, scale-free connection length distribution and so on. For a detail review of the field, see~\cite{PR1}.

The most important features of spatial network model is that there is a cost associated with the length of links, which has dramatic effects on the topology and function of these networks. A total cost constraint has been recently introduced to design the spatial networks~\cite{PRL104018701,PRL106108701,EPL8958002,EPL9258002,PhysA3903962,PRE81025202}. The total cost $C$ is defined as the total length of the links, $C=\sum r$ where $r$ is the length of links. In \cite{PRL104018701}, pairs of sites $ij$ in $2$-dimensional lattices are randomly chosen to be linked with long-range connections with probability $P_{ij}\sim r_{ij}^{-\alpha}$, where $r_{ij}$ is the Manhattan distance between sites $i$ and $j$. New links are added until the total length of the links reach the total cost $C$. The exponent $\alpha$ controls the trade-off between the link length and link number. A large value of $\alpha$ allows for the formation of many short-length links while the small $\alpha$ favors the creation of a few long-length links. The authors in \cite{PRL104018701} show that the optimal exponents for both the average shortest-path length and navigation steps are $\alpha=3$ in 2-dimension and $\alpha=2$ in 1-dimension (the 1-dimensional scenario has been strictly proved very recently in \cite{EPL9258002}). The authors claim that such model reveals the optimized aspect of airline networks under the conditions of geographical availability (for customer satisfaction) and cost limitations (for airline company profit). Similar works have been carried out to study dynamics such as traffic congestion and synchronization on spatial networks under total cost constraint, and the optimal link length exponents are found~\cite{EPL8958002,PhysA3903962,PRE81025202}.

However, we observe that the degree distribution in these spatial network models with total cost constraint is homogenous, which is far different from real cases, including airline systems. Many previous works have revealed that many geographic-based networks have heterogenous degree distribution~\cite{PNAS7794}. In this paper, we propose a method to introduce degree heterogeneity in spatial networks with total cost constraint. We find that degree heterogeneity affects the optimal exponent for the link length distribution. Specifically, the exponent shifts to a smaller value. Also, the average shortest-path length can further decrease due to degree heterogeneity. Moreover, we study the effect of degree heterogeneity on the dynamics taking place on the spatial networks, and we observe the same shifting phenomenon of the optimal exponent.

To begin our analysis, we first briefly describe the original spatial network model introduced in \cite{PRL104018701}. Nodes are located in a $d$-dimensional regular square lattice, where each site $i$ is connected
with its $2d$ nearest neighbors. Pairs of sites $ij$ are randomly chosen to receive long-range connections with probability proportional to $r_{ij}^{-\alpha}$, where $r_{ij}$ is the Manhattan distance between sites $i$ and $j$ (i.e., the number of connections separating the nodes in the underlying regular lattice). Long-range connections are added to the system until their total length (cost) $\sum r_{ij}$ reaches a given value $C$. The exponent $\alpha$ controls the average length of the long-range connections, a smaller $\alpha$ is corresponding to fewer but longer connections, and vice versa. In this model, the most interesting phenomenon is that the average shortest-path length and the navigation steps of the spatial networks can be minimized under a certain value of the parameter $\alpha$. Specifically, optimal values are found to be $\alpha^{*}=3$ in $2$-dimensional spatial networks and $\alpha^{*}=2$ in $1$-dimensional ones. Accordingly, it is concluded that the best transportation condition is obtained with an exponent $\alpha^{*}=d+1$ where $d$ is the dimension of the underlying lattice.

\begin{figure}
  \center
  \includegraphics[width=9cm]{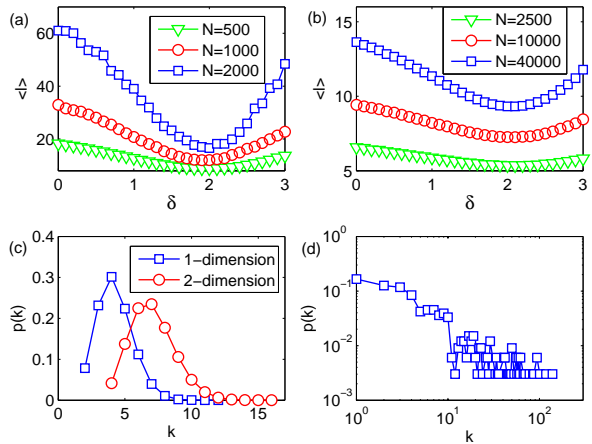}
\caption{(Color online) The average shortest-path length $\langle l \rangle $ as a function of $\delta$ in spatial networks~\cite{PRL104018701} embedded in (a) $1$-dimensional lattices and (b) $2$-dimensional lattices. (c) and (d) are respectively the degree distribution of these network models ($N=1000$ for $1$-dimension and $N=10000$ for $2$-dimension, $\delta=2$) and the degree distribution of the airline network of USA. Results for the $1$-dimensional and $2$-dimensional models are averaged over $100$ and $10$ independent realizations, respectively.}
\label{fig1}
\end{figure}

The probability $P_{ij}$ for two nodes ($i$ and $j$) to receive a long-range
connection can be described by the probability density function (PDF)
$p(r)$ of long-range connections with Manhattan length $r$. Actually, $p(r)$ is the summation of $P_{ij}$ with $r_{ij}=r$. In $d$-dimensional space, the number of nodes which have Manhattan distance $r$ from a given
node is proportional to $r^{d-1}$. Therefore, $p(r)$ can be expressed as
\begin{equation}
p(r)\sim r^{d-1}r^{-\alpha}=r^{d-\alpha-1},
\end{equation}
which means $p(r)\sim r^{-\delta}$ and $\delta=\alpha+1-d$. Since the optimal exponent is $\alpha^{*}=d+1$, $\delta^{*}=(d+1)+1-d=2$ independently of spaces' dimension.

Based on this observation, the spatial network in \cite{PRL104018701} can be equivalently generated by the following procedure:
\begin{enumerate}
\item $N$ nodes are arranged in a $d$-dimensional lattice with periodic boundary condition. Every node is connected with its nearest neighbors, which
can keep every node reachable. In addition, between any pair of
nodes there is a well defined Manhattan distance.
\item A certain distance
$r$ $(2\leq r\leq r_{\rm max}$, where $r_{\rm max}$ is the largest distance between
any nodes in the initial network) is generated with probability
$P(r)=ar^{-\delta}$, where $a$ is determined from the normalization
condition $\sum^{r_{\rm max}}_{r=2}P(r)=1$.
\item A node $i$ is chosen randomly, and one of the $N_{r}$ nodes who are at distance
$r$ from node $i$ is picked randomly, and an edge
between them is created if there is no edge between
them yet.
\item  Update the total cost $\sum r$. Repeat step 2 and 3 until the total
cost reaches $C=cN$ (In the following, we fix $c=10$ for convenience).
\end{enumerate}

We first study the average shortest-path length $\langle l \rangle$ in spatial networks generated by the above procedure (see fig. 1(a) and (b)). As expected, the optimal $\langle l \rangle $ appears under the exponent $\delta=2$ in both $1$-dimension and $2$-dimension networks, and this feature is independent of the network size. In addition, we observe that the degree in such spatial networks is homogenously distributed, as shown in fig. 1(c). However, the degree distribution in real system is far more heterogeneous~\cite{PR1}. For example, the Chinese airline~\cite{PRE69046106}, the Italian airline~\cite{CSF31527} and the world wide airline~\cite{PNAS1027794} networks are found to have truncated power-law degree distribution. We show in fig. 1(d) that for the airline network of USA (USAir)~\cite{USAirdata} the degree distribution approximately obeys a power-law. To capture this feature of real networks, it is necessary to introduce degree heterogeneity in spatial network models and investigate how the topology and function of the networks are affected. Accordingly, we propose a revised spatial network generating procedure. Instead of randomly adding the long connection to the underlying lattice in step $3$, we first calculate a score $S_{ij}$ for each pair of node $ij$ with distance $r$ (generated in step $2$) and add a link between the pair $ij$ with probability proportional to its score. $S_{ij}$ is obtained by
\begin{equation}
S_{ij}=(k_{i}k_{j})^{\theta},
\end{equation}
where $\theta$ is a tuneable parameter. When $\theta=0$, the generated network reduces to the original model in \cite{PRL104018701}. When $\theta<0$, new links will connect the nodes with lower degree, which leads to an even more homogeneous degree distribution. When $\theta>0$, new links will preferentially attach to the nodes with higher degree, and a heterogenous degree distribution emerges. Two examples are shown in fig. 2(a) and (b). Obviously, by adjusting the parameter $\theta$, the degree distribution is no longer Poissonian as in the original model. When $\theta$ is small (e.g. $\theta=2$), the spatial networks exhibit a power-law-like degree distribution. In this case, the network has several sub hubs which are connected to each other by some long distance links. When $\theta$ is relatively large (e.g. $\theta=5$), a few super hubs will emerge and most of the links are connected to such nodes. Therefore, the tail of the degree distribution will become longer and longer as $\theta$ increases. The degree heterogeneity can be measured by the index $H=\langle k^2 \rangle /\langle k \rangle ^2$. The higher the $H$, the larger the degree heterogeneity. As shown in fig. 2(c), the index $H$ increases with $\theta$, which confirms that the parameter $\theta$ can indeed control the degree heterogeneity. In the USAir network, $H=3.46$. It suggests that the degree heterogeneity of real system is far from that of the original network model where $\theta=0$ and $H \thickapprox 1$. Furthermore, the modified spatial model can perfectly preserve the link length distribution as the original model since it doesn't affect the link length generating process but only rearranges the location of the links. We can see from fig. 2(d) that the link length distributions for different $\theta$ is identical as for the case $\theta=0$.

\begin{figure}
  \center
  \includegraphics[width=9cm]{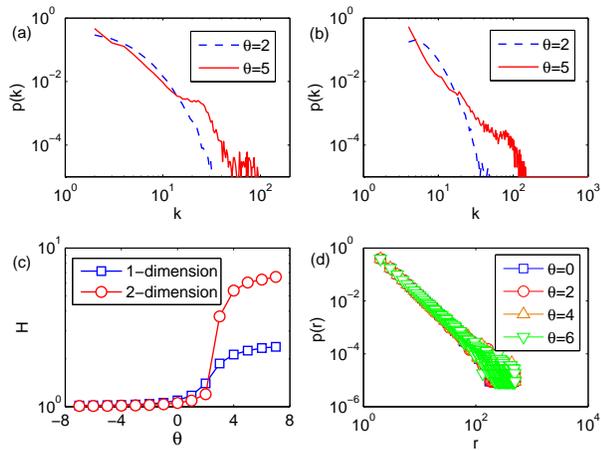}
\caption{(Color online) The degree distribution of the (a) $1$-dimensional and (b) $2$-dimensional modified spatial network model under different $\theta$.
(c) shows the index $H$ as a function of the parameter $\theta$ and (d) is the link length distributions for different $\theta$ in $1$-dimensional spatial networks. The parameters are set as $\delta=2$, $N=1000$ for $1$-dimension and $N=10000$ for $2$-dimension. Results for the $1$-dimensional and $2$-dimensional models are averaged over $100$ and $10$ independent realizations, respectively.}
\label{fig2}
\end{figure}

Since the average shortest-path length $\langle l \rangle$ reflects the general transportational ability for a network, we conduct simulations for different values of $\theta$ and see how the degree heterogeneity affects the optimizing process of $\langle l \rangle$ based on different $\delta$. The simulations are carried out in both $1$-dimensional and $2$-dimensional spatial models and the results are reported in fig. 3. When $\theta\leq0$, the minimal $\langle l \rangle $ is always achieved at $\delta^{*}=2$. However, when the degree distribution becomes heterogenous (i.e., $\theta>0$), the optimal exponent $\delta^{*}$ will shift to a smaller value. The results are similar in both $1$-dimensional and $2$-dimensional spatial networks (see fig. 3(a) and (b) respectively), and imply that a few more longer connections are needed to minimize $\langle l \rangle$ when network degree distribution is heterogenous.

\begin{figure}
  \center
  \includegraphics[width=9cm]{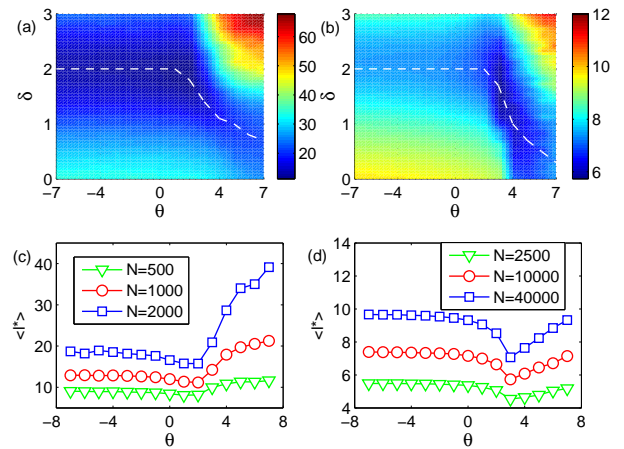}
\caption{(Color online) The dependence of $\langle l \rangle $ on $\delta$ under different $\theta$ in spatial networks generated from (a) $1$-dimensional lattices with $N=1000$ and (b) $2$-dimensional lattices with $N=10^{4}$. The white lines mark the optimal $\delta^{*}$ under different $\theta$. (c) and (d) are the relation between $\theta$ and the minimal $\langle l^{*} \rangle $ in $1$-dimensional and $2$-dimensional spatial networks respectively. The results for the $1$-dimensional and $2$-dimensional models are averaged over $100$ and $10$ independent realizations respectively.}
\label{fig3}
\end{figure}

For each $\theta$ there is an optimal $\delta^{*}$ and a minimal $\langle l^{*} \rangle $, another interesting aspect is hence studying the relation between $\theta$ and $\langle l^{*} \rangle $. As we can see from fig. 3(c) and (d), introducing the degree heterogeneity can indeed further decrease $\langle l^{*} \rangle $. However, if one keeps enlarging $\theta$ , $\langle l^{*} \rangle $ will start to increase. We can see from fig. 3(c) and (d) that optimal $\theta^{*}$ for $\langle l^{*} \rangle $ exist in both $1$-dimensional and $2$-dimensional spatial networks. Interestingly, these optimal $\theta^{*}$ is independent of the network size but dependent on the network dimension. Specifically, $\theta^{*}=2$ in $1$-dimensional spatial networks and $\theta^{*}=3$ in $2$-dimensional spatial networks, which leads us to conjecture that the optimal value is obtained at $\theta^{*}=1+d$. We also consider the greedy navigation algorithm used in ref.~\cite{PRL104018701} to model the case when only local information of the network is known during the transport process. Similar optimal exponent shifting phenomenon is observed and the minimum navigation steps can further decrease due to the degree heterogeneity.

Another important aspect to consider in spatial networks is how the network function (dynamics) can be enhanced. Specifically, refs.~\cite{PhysA3903962,PRE81025202} have revealed that there is an optimal link length distribution for the network synchronizability. Under the framework of master stability analysis, the synchronizability of an undirected network can be quantified by the eigenvalue ratio of the corresponding Laplacian matrix of this network, namely $R=\lambda_{N}/\lambda_{2}$, where
$\lambda_N$ and $\lambda_2$ are respectively the largest and the smallest non-zero eigenvalues of the Laplacian matrix~\cite{PR93}. Generally, the smaller the value of $R$, the stronger the network synchronizability. In \cite{PhysA3903962,PRE81025202}, starting from $1$-dimensional lattice, the authors find the optimal length distribution of links (power-law distribution with exponent $\delta=1.5$) to add to the network for minimizing $R$. However, these previous network models also yields homogeneous degree distribution. Here we are interested in how the degree heterogeneity affects the synchronizability for spatial networks. The results are reported in fig. 4. Obviously, the optimal $\delta^{*}$ for $R$ shifts from $\delta^{*}=1.5$ to a smaller value as $\theta$ increases. However, the optimal synchronizability $R^{*}$ cannot be further enhanced by introducing the degree heterogeneity as shown in fig. 4(b). It is well known that degree heterogeneity will weaken the synhcornizability while a small average shortest-path length is generally favorable for synchornizability~\cite{PR93}. As we discussed above, introducing the degree heterogeneity can further decrease the average shortest path. However, the decreased average shortest-path length in current case is not sufficient to overcome the effect of degree heterogeneity, and hence $R^{*}$ keeps increasing with $\theta$.

\begin{figure}
  \center
  \includegraphics[width=9cm]{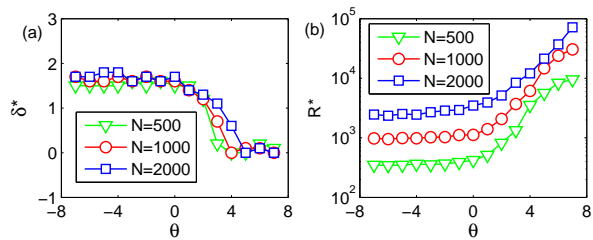}
\caption{(Color online) (a) The dependence of the optimal $\delta^{*}$ for synchronizability $R$ on $\theta$. (b) The optimal synchronizability $R^{*}$ as a function of $\theta$. The spatial networks are built on $1$-dimensional lattices. The results are averaged over $100$ independent realizations.}
\label{fig4}
\vspace{-0.3cm}
\end{figure}

To summarize, we have proposed a method to introduce degree heterogeneity in spatial networks without destroying the power-law link length distribution of the original model~\cite{PRL104018701}. We have investigated its effect on network topology and function. When optimizing the average shortest-path length and the network synchronizability, we find that the best power-law exponent for the link length distribution shifts to a smaller value than that of the original model. Additionally, the average shortest-path length can further decrease due to the degree heterogeneity while the network synchronizability is weakened if the degree distribution is heterogenous.

From the practical point of view, the original spatial network model~\cite{PRL104018701} can explain the power-law connection length distribution with given exponent in some real systems. By building the relation between the optimal exponent and the degree heterogeneity, our model can extend such explanation of real transportation networks to a wider range of connection length distribution which is found to be existing by many empirical studies~\cite{PR1}. Furthermore, since we reveal the optimal degree heterogeneity for average shortest path, our model may explain why the degree heterogeneity is widely exhibited in real systems and only within certain range of values. In this sense, our model can not only better understand and model real transportation systems, but also help to design efficient ones.

We would like to thank Yi-Cheng Zhang, Giulio Cimini, Matus Medo and Chi Ho Yeung for helpful suggestions. This work is supported by the Swiss National Science Foundation (No. 200020-132253).

\end{document}